\documentclass[twocolumn,preprintnumbers,amsmath,amssymb,superscriptaddress,showpacs]{revtex4}

\usepackage{graphicx}
\usepackage{color}
\usepackage{pslatex}
\usepackage{dcolumn}
\usepackage{bm}
\begin{document}


\title{Thermal and mechanical properties of a DNA model with solvation barrier}

\author{R. Tapia-Rojo}

\affiliation{Dpto. de F\'{\i}sica de la Materia Condensada,
Universidad de Zaragoza. 50009 Zaragoza, Spain}

\affiliation{Instituto de Biocomputaci\'on y F\'{\i}sica de Sistemas
Complejos, Universidad de Zaragoza. 50009 Zaragoza, Spain}

\author{J.~J. Mazo}

\affiliation{Dpto. de F\'{\i}sica de la Materia Condensada,
Universidad de Zaragoza. 50009 Zaragoza, Spain}

\affiliation{Instituto de Ciencia de Materiales de Arag\'on,
C.S.I.C.-Universidad de Zaragoza. 50009 Zaragoza, Spain}

\author{F. Falo}

\affiliation{Dpto. de F\'{\i}sica de la Materia Condensada,
Universidad de Zaragoza. 50009 Zaragoza, Spain}

\affiliation{Instituto de Biocomputaci\'on y F\'{\i}sica de Sistemas
Complejos, Universidad de Zaragoza. 50009 Zaragoza, Spain}

\date{\today}

\begin{abstract}

We study the thermal and mechanical behavior of DNA denaturation in
the frame of the mesoscopic Peyrard-Bishop-Dauxois model with the
inclusion of solvent interaction.  By analysing the melting transition
of a homogeneous A-T sequence, we are able to set suitable values of
the parameters of the model and study the formation and stability of
bubbles in the system. Then, we focus on the case of the P5 promoter
sequence and use the Principal Component Analysis of the trajectories
to extract the main information on the dynamical behaviour of the
system. We find that this analysis method gives an excellent agreement
with previous biological results.

\end{abstract}

\maketitle

\section{Introduction}

In the last years there exist an increasing interest in the
description of complex biological systems using simple physical
models~\cite{PhysicsMolBio}. Physical models for biological problems
should contain the key ingredients to explain their basic
phenomenology . In this sense, DNA melting can be modeled using very
simple statistical mechanics models~\cite{PhysRep}. One of the most
successful is the one dimensional Peyrard-Bishop-Dauxois (PBD)
model~\cite{PBD,Peyrard_NonL} which with very few assumptions is able
to reproduce the melting curves for different DNA sequences. PBD model
undergoes an entropic phase transition between the native closed state
and a denatured one in which the DNA strands are separated. The nature
of this transition strongly depends on model parameter, so different
versions of the model have been proposed in the last years.

A virtue of the PBD model is that it allows to study not only the
equilibrium properties of the molecule but also dynamical ones; for
instance, the formation and stability of the so called DNA bubbles
(short open segments of the DNA chain)~\cite{Peyrard_Nature,
  Peyrard_JPCM}. Due to its versatility, this simple model has been
applied in other contexts beyond the study of the equilibrium melting
curves. Thus, it has been used to investigate the DNA mechanical
denaturation~\cite{SCL_JBP, Voulgarakis} and it has been specially
successful in the modelling of the open regions in short DNA
hairpins~\cite{Peyrard_JPCM,Ares_PRL, Ares_thesis}. This fact has
motivated to extend the model to study non-homogeneous sequences and
to try to correlate the \emph{dynamical} behavior of the model to
\emph{functional} aspects of the DNA chain. To be specific, it has
been suggested that the localised dynamical excitations (bubbles) of
the model are directly related with the protein-DNA binding
sites~\cite{Choi, Kalosakas04}.  This point has been controversial,
due to the simplicity of the PBD model assumptions~\cite{Erp1, Erp2}.

One of the most interesting improvements to the model is the inclusion
of a barrier in the intra-base interaction term~\cite{Weber}. Such
barrier accounts for solvent interactions in the system. The addition
of barriers modifies deeply the nature of the denaturation transition
and the dynamics of the molecule. Depending on the parameter values,
the transition become sharper, even in the harmonic stacking energy
version. The transition width has not a trivial behavior and it
deserves a more careful study. On the dynamical side, the bubbles
lifetime increases dramatically with the inclusion of this term and
approaches to the experimental observations.

The purpose of this paper is twofold. First, we characterize the
melting transition of homogeneous sequences and the formation of
bubbles phenomenon for a wide range of parameters (section IV).  This
allows us to set suitable parameter values for next research.  Then,
we focus on the problem of the relation between dynamics and function.
In section V, we study a heterogeneous sequence which contains known
information on the transcription process of DNA to mRNA (a P5 virus
promoter). We use Principal Component Analysis of trajectories to
identify the regions in the sequence which most relevantly contribute
to the dynamical fluctuations of the molecule. These regions are the
softest ones, from a mechanical point of view, and they correlate
fairly well with the relevant biological sites. Given its simplicity
and efficiency, we propose to extend this analysis to other sequences
in order to get insight over functionality of the different genome
regions.

\section{THE MODEL}

We study a modification of the Peyrard-Bishop-Dauxois model. In the
PBD model the complexity of DNA is reduced to the study of the
dynamics of the $N$ base pairs of the molecule. For each base pair we
define the variable $y_n$ associated to the distance between the
bases. The total energy of the system is then approached by:
\begin{equation}
H=\sum_{n=1}^N \left[ \frac{p_n^2}{2m}+V(y_n)+W(y_n,y_{n-1}) \right].
\label{eq:ham}
\end{equation}
Here $p_n=m \; dy_n/dt$, $n$ is the index of a base pair and $m$ its
reduced mass. 

In this equation, we identify two energy potential terms: $V(y_n)$, an
on-site potential one, and $W(y_n, y_{n-1})$ which accounts for
inter-pairs interactions. The standard PBD model corresponds to a
particular choice for these energy terms.

\begin{figure}[]
\centering{\includegraphics[width=0.5\textwidth]{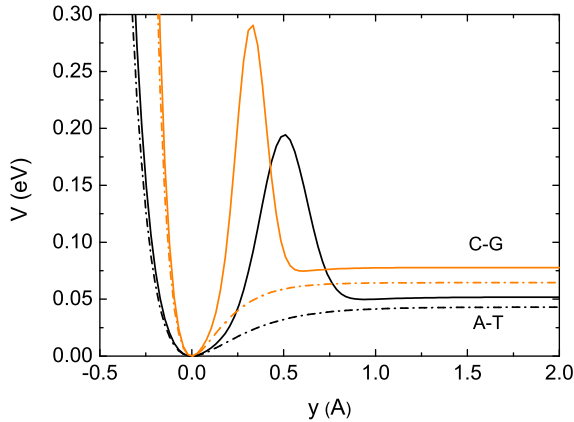}}
\caption{(Color online) Morse potential with (solid lines) and without (dashed lines)
  solvation barrier for A-T (black) and C-G (orange/light gray) base pairs. For A-T
  pair without barrier $\alpha=4 {\rm \AA}^{-1}$ and $D=0.043$eV. For
  A-T with barrier $\alpha=4 {\rm \AA}^{-1}$, $D=0.05185$eV, $G=3D$,
  $y_0=2/\alpha$ and $b=1/2\alpha^2$ (the value of $D$ has been
  adjusted in each case to give the desired melting temperature for
  the A-T uniform chain). For C-G pairs $D_{\rm CG}=1.5 D_{\rm AT}$
  and $\alpha_{\rm CG}=1.5 \alpha_{\rm AT}$.}
\label{fig:V} 
\end{figure}

The potential $V(y_n)$ describes the interaction between the two bases
of a pair. The PBD model uses a Morse potential to account for such
interaction (see Fig.~\ref{fig:V}). This is an standard approximation
for chemical bonds.
\begin{equation}
V(y)=D ({\rm e}^{-\alpha y}-1)^2,
\label{eq:V}
\end{equation}
where $D$ corresponds to the dissociation energy of the pair [$V(0)=0$
  and $V(\infty)=D$] and $\alpha$ sets the amplitude of the potential
well [$V''(0)=2D\alpha^2$]. See also that variable $y$ measures
deviations of the base distance with respect to equilibrium, thus
$V(y)$ has a minimum for $y=0$.

An important contribution to the stability of DNA molecule comes from
the stacking of the bases. The potential $W(y_n, y_{n-1})$ describes
the interaction between base pairs along the DNA strand. The simplest
approximation is to consider just harmonic interaction between
adjacent pairs in the molecule. However this choice does not reproduce
the observation of a sharp thermal denaturation curve. To solve this,
the PBD model includes a position dependent nonlinear coupling
constant
\begin{equation}
W(y_n, y_{n-1})=\frac{1}{2} K (1+\rho \; {\rm
  e}^{-\delta(y_n+y_{n-1})})\; (y_n-y_{n-1})^2.
\label{eq:W}
\end{equation}
The effect of this term, whose intensity is governed by $\rho$ is to
change the effective coupling constant from $K(1+\rho)$ to $K$ when
one of the base pairs is displaced away from its equilibrium
position. The parameter $\delta$ sets the scale length for this
behavior.

DNA melting or denaturation refers to the separation of the two
strands of the DNA molecule to generate two single strands.  The great
merit of the PDB model defined in Eqs.~\eqref{eq:V} and~\eqref{eq:W} is
to reproduce the different melting curves observed in DNA
molecules. Thus, it has been used as a solid starting point for many
other statistical mechanics studies of DNA.

The melting process involves the breaking of hydrogen bonds between
the bases. In order to study different DNA sequences the PBD model
also can include sequence dependent Morse potential parameters: $D_n$
and $\alpha_n$. It can be intuitively deduced from the fact that A-T
pairs are linked by two hydrogen bonds meanwhile C-G ones are linked
by three, thus forming a more stable link. Following~\cite{campa} in our
simulations we will use $D_{\rm CG}=1.5 D_{\rm AT}$ and
$\alpha_{\rm CG}=1.5 \alpha_{\rm AT}$. Although recent works have
considered more complex sequence dependence~\cite{Peyrard_JPCM,Choi2}, we will not
introduce more complexity at this level in the model.

We will work with a modified version of the PBD model to include
solvent interaction. Solvent interaction stabilizes open pair states
by means of the hydrogen bonds that the base pairs may form with the
solvent when opened~\cite{Weber}. Such bonds have to be broken before
the pair closes again. This effect can be included in a simple way
with an effective barrier in the Morse potential.  The addition of the
barrier has to avoid, as far as possible, any other effect in the
Morse well. Hence, we have chosen the following definition for the
on-site potential $V(y)$:
\begin{equation}
V(y)=D ({\rm e}^{-\alpha y}-1)^2 + G {\rm e}^{-(y-y_0)^2/b}.
\label{eq:Vmod}
\end{equation}
This potential results from the addition of a gaussian barrier, whose
height is controlled by $G$, the position is given by $y_0$ and its
width by $b$. A reasonable election for such parameters is $G=3D$,
$y_0=2/\alpha$ and $b=1/2\alpha^2$. Figure~\ref{fig:V} plots the
intra-base potential $V$ for an A-T and a C-G base pair.

\section{METHODS}

\subsection{Langevin Dynamic Simulations.}
 
In order to study the behavior of the system we have performed
molecular dynamics numerical simulations of the Langevin equation:
\begin{align}
 m \frac{d^2 y_n}{dt^2}  & +  m \gamma \frac{d y_n}{dt} + 
 \frac{\partial  V(y_n)}{\partial y_n} + \nonumber \\
& + \frac{\partial  [W(y_n,y_{n-1})+W(y_{n+1},y_n)]}{\partial y_n} = \xi_n(t),
\label{eq:lang_dyn}
\end{align}
where $m$ is the mass of the pair, $\gamma$ the effective damping of
the system and $\xi(t)$ accounts for thermal noise, $\langle \xi_n (t)
\rangle=0$ and $\langle \xi_n (t) \xi_k (t') \rangle= 2 m \gamma k_B T
\delta_{nk} \delta (t-t')$ with $T$ the bath temperature.

The equations were numerically integrated using the stochastic
Runge-Kutta algorithm~\cite{sde1,sde2}. Simulation of large A-T chain
(section IV) used periodic boundary conditions in order to avoid any
terminal effect. For the P5 promoter (section V) we used fixed
boundary conditions which will be described below.

To best characterize the phase transition we have computed the mean
energy $\langle u \rangle$, and mean displacement $\langle y \rangle$ defined as
\begin{equation}
\langle u \rangle = \frac{1}{N t_s} \; \sum_{n,t}^{N,t_S} \left[ W(y_n,y_{n-1}) + V(y_n)
\right]
\label{eq:energy}
\end{equation}
and
\begin{equation}
\langle y \rangle = \frac{1}{N} \sum_{n=1}^{N}\langle y_n \rangle = \frac{1}{N t_s} \; \sum_{n,t}^{N,t_S} y_n (t).
\label{eq:md}
\end{equation}
$N$ is the number of base pairs in the sequence to study and $t_s$ the
total simulation time.

We are also interested in computing the probability of being opened
for each base. We define this as follows:
\begin{equation}
P_n(y_{\rm th})=
\frac{1}{t_s} \;
\sum\limits_t^{t_s} \Theta (y_n(t)-y_{\rm th})
\label{eq:prob}
\end{equation}
where $y_{\rm th}$ is a threshold for opening that we choose just
behind the barrier of the Morse potential and $\Theta(x)$ is the
Heaviside step function ($\Theta(x)=0$ for $x<0$ and $\Theta(x)=1$ for
$x\geq0$).  Thus $P_n$ indicates the fraction of time the $n$-th base
is opened during the time $t_s$.

\subsection{Principal Component Analysis (PCA)}

PCA is a statistical method to extract information from a large set of
data in a multidimensional phase space, allowing to reduce the
dimensionality of the variables to those that include most of the
fluctuations of the original system~\cite{jolliffe_book}.  This is
achieved by a change of coordinates in the phase space and the new
coordinates are the so-called Principal Components, PCs. This method has become
a standard tool in the analysis of trajectories in "all-atom" simulations of 
macromolecules \cite{amadei}.

Operationally, we have to build the $N \times N$ correlation matrix
\begin{equation}
 C(i,j) = \left< y_i y_j \right> - \left< y_i \right> \left< y_j \right>.
\label{eq:correl}
\end{equation}
Diagonalising this matrix we obtain an ordered set of eigenvalues
($\lambda_1 > \lambda_2 > \lambda_3 ...$) with their corresponding
eigenvectors ($v_1, v_2, v_3 ....$).  Eigenvalue $\lambda_i$ gives the
amount of fluctuations which corresponds to eigenvector $v_i$, thus
the new coordinates are ordered in such a way that the few first ones
retains most of the fluctuations of the system. It is useful to define
a frequency $\omega_i$ associated to each eigenvalue $\lambda_i$ and
given by
\begin{equation}
 \omega_i = \sqrt{\frac{k_B T}{\lambda_i}}.
\label{eq:freq}
\end{equation}
From this point of view, the larger fluctuations will correspond to
the lowest frequencies. It can be proved that in the low temperature
limit these frequencies converge to the normal mode frequencies of
the system. Normal mode analysis has also used in the detection of relevant
motions in coarse-grained model of proteins \cite{bahar}.

\section{The uniform chain}

\subsection{Fitting the phase transition: parameters of the model}

\begin{figure}[]
\centering{\includegraphics[width=0.5\textwidth]{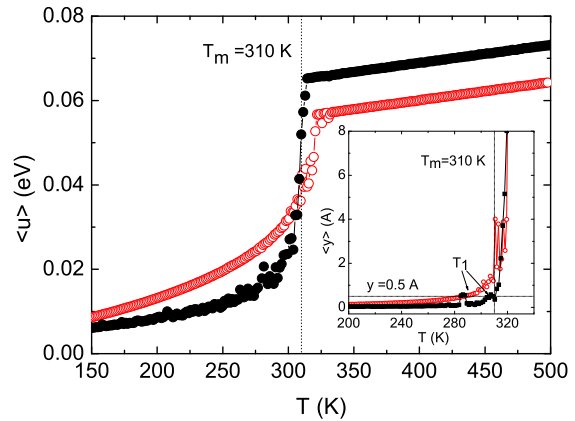}}
\caption{(Color online) The melting transition. Mean energy and displacement (inset)
  as a function of temperature of a homogeneous chain for the
  standard (open circles) PBD model and the modified one (full circles). Parameters
  are given in the text.}
\label{fig:melting} 
\end{figure}

In this section we will study the melting transition of a homogeneous
chain of A-T pair bases.  Although there are some studies on the
influence of the different parameters in the transition, up to our
knowledge, there is not a systematic scan over the parameter space. In
particular, when the barrier is included on the Morse potential a
clear narrowing of the transition is observed (see
Fig.~\ref{fig:melting}). The same effect is expected when the $\rho$
parameter is increased.  Thus, one of the first problems when dealing
with this modified PBD model is to find suitable model
parameters. This is mainly done by adjusting the theoretical melting
temperatures to the experimental ones. The melting temperature $T_m$
is usually defined as the temperature at which the DNA strands are
half denatured. For a homogeneous A-T chain, the melting transition
has been reported around $T_{m} = 310 K $~\cite{Choi2, wells}.

One should be aware that, in general, the melting temperature depends
on the length of the sequence, base composition, topological structure
and salt concentration. Thus, it is difficult to predict the exact
transition temperature of a given sequence. In the case of the uniform
chain the melting temperature depends strongly on salt concentration,
which is not explicitly included in this mesoscopic model. Usual
parameters for the standard PBD model can be found in
references~\cite{campa,Choi2} for instance. Following previous
findings and guided by our numerical simulation results we choose the
following set of values for studying the standard model: $m=300$ Da, $D=0.043$eV,
$\alpha=4 {\rm \AA}^{-1}$, $K=0.03$eV$\times {\rm \AA}^{2}$, $\rho=3$
and $\delta=0.8 {\rm \AA}^{-1}$ (these values are slightly different
to those given in~\cite{Choi2}). If barrier is included we modify the
value of $D$ to obtain the same $T_m$. Thus, in this case
$D=0.05185$eV, and $G=3D=0.1556$eV, $y_0=2/\alpha=0.5 {\rm \AA}$ and
$b=1/2\alpha^2=0.03125 {\rm \AA}^2$.

The melting transition curve for the model with and without solvation
barrier is shown in Fig.~\ref{fig:melting} for a homogeneous A-T
sequence with $N=220$. At high temperature is observed the expected
linear behavior of a free gaussian polymer chain; i.e. the completely
unzipped state.

\begin{figure}[t]
\centering{\includegraphics[width=0.5\textwidth]{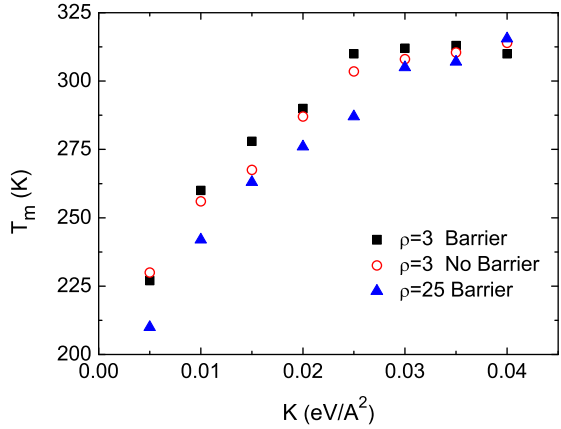}}
\caption{(Color online) Transition temperature versus the stacking
  constant $K$ for various $\rho$ values.}
\label{fig:Tm} 
\end{figure}

\begin{figure}[]
\centering{\includegraphics[width=0.5\textwidth]{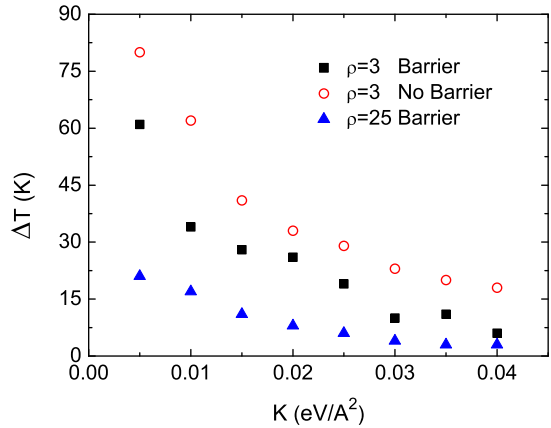}}
\caption{(Color online) Transition width versus the stacking constant
  $K$ for various $\rho$ values.}
\label{fig:dT} 
\end{figure}

\begin{figure}[]
\centering{\includegraphics[width=0.5\textwidth]{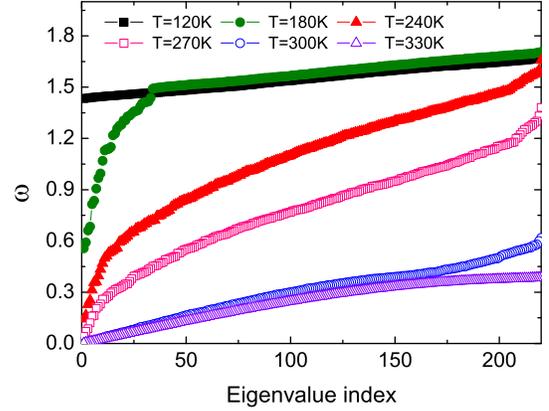}}
\caption{(Color online) PC frequency spectrum at different
  temperatures between 120 and 330K. Frequency units are $(D/m)^{1/2}\alpha = 5.15\times10^{12} \rm sec^{-1}$.}
\label{fig:PCAfreq} 
\end{figure}

\begin{figure}[]
\centering{\includegraphics[width=0.5\textwidth]{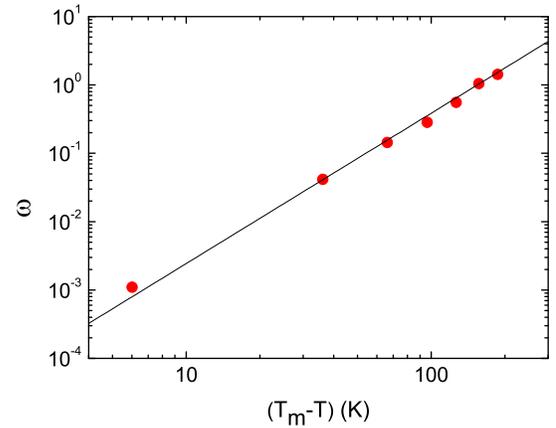}}
\caption{(Color online) Temperature dependence of the lowest PC frequency. The
  straight line corresponds to the best fit with the parameters given
  in the text. Frequency units are $(D/m)^{1/2}\alpha = 5.15\times10^{12} \rm sec^{-1}$.}
\label{fig:zeroM} 
\end{figure}

The numerical determination of the melting temperature is not a
trivial issue. For the case of a uniform chain, we use the following
computational criteria to determine the melting temperature and
transition width from the $\langle u(T) \rangle$ and $\langle y(T)
\rangle$ curves. First we define two temperatures. The larger one,
$T_2$, gives the onset of the linear behavior in $\langle u(T)
\rangle$ which indicates that the chain is completely melted. The
other one, $T_1$ estimates the beginning of the transition defined as
the temperature for which $\langle y(T) \rangle = y_0$,
i.e. when the chain can be considered to be by average in the barrier
position. Then we define the transition width $\Delta T = T_2 -
T_1$ and the melting temperature as $T_m = (T_1+T_2)/2$.

It is interesting to briefly discuss on the effect of the main model
parameters in the melting transition and the dynamical behavior of the
system. High values of $D$, the Morse potential dissociation energy,
produces also high $T_m$.  Parameters $K$ and $\rho$, set the stacking
interaction between adjacent bases. Its effect in $T_m$ and $\Delta T$
can be seen in Figs.~\ref{fig:Tm} and~\ref{fig:dT}. A suitable melting
temperature and transition width is obtained at high $K$ values (thus
we chose $K=0.03$eV/${\rm \AA}^2$) and moderate $\rho$ ($\rho \sim
3$). Moderate $\rho$ are biologically satisfactory and avoid the very
narrow bubbles observed for high $\rho$ values. With respect to the
solvation barrier its presence makes bubbles live longer. Thus a
complete separation of the strands is also facilitated leading to a
decrease of the melting temperature (this change can be
counterbalanced by an increasing in the $D$ value). The inclusion of
the barrier also reduces the width of the transition, allowing the
election of moderate $\rho$ values.

\begin{figure*}[]
\centering{\includegraphics[width=0.7\textwidth]{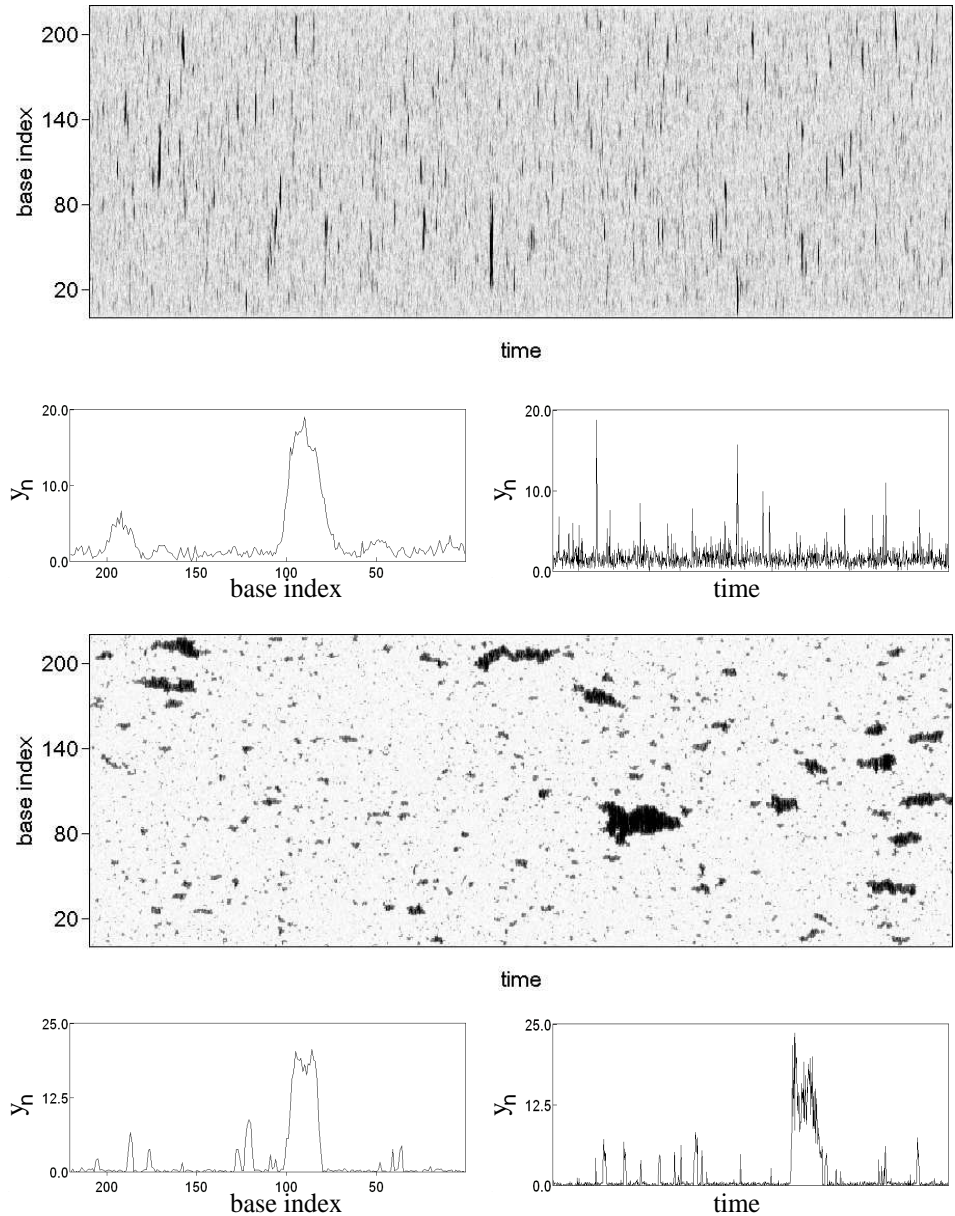}}
\caption{Typical molecular dynamics simulation trajectories for the
  PBD model of a homogeneous sequence without barrier (upper figures)
  and with barrier (lower ones). Dark areas correspond to open base
  pairs. Long-living bubbles are clearly observed when solvation
  barrier is included. Small figures show a base pair configuration at
  a given time (left) and the trajectory of a given base as a function
  of time (right). Trajectory time is 200 ns and $y_n$ is given in
  units of $\alpha^{-1} = 0.25 \AA$. }
\label{fig:bubbles} 
\end{figure*}

\subsection{PCA of the phase transition}

PCA provide us another method to analyse the phase
transition. Figure~\ref{fig:PCAfreq} shows the spectrum of effective
frequencies as defined in Eq.~\eqref{eq:freq} for the uniform A-T chain.
At low temperatures, a sharp band of linear waves excitations close to
the ground state is clearly identified. At very low temperatures, the
frequencies are given by the dispersion relation~\cite{comm1}:
\begin{equation}
 m\omega^2 \simeq 2D \alpha^2 + 2K(1 + \rho) [1- \cos( \pi n/N)].
\label{eq:dispersion}
\end{equation}
Thus, the gap for the lowest mode is controlled by the intra-base
potential parameters. In this limit the PC eigenvectors are the
normal modes of a homogeneous array. As temperature increases, the
nonlinear excitations are more important. Such nonlinear modes are
responsible of the larger fluctuations and correspond to the larger
eigenvalues or lower frequencies. As we approach the melting
temperature a soft mode goes to zero (see Fig.~\ref{fig:PCAfreq}),
the chain is not pinned by an on-site potential and every strand moves
freely. Hence at high temperature the frequencies are those of a free
gaussian chain with coupling given by the stacking parameter $K$:
\begin{equation}
 m\omega^2 =  2K [1- \cos( \pi n/N)].
\label{eq:dispersion2}
\end{equation}
Note that the zero mode corresponds to the largest wavelength.

Figure~\ref{fig:zeroM} shows the temperature evolution of this mode
frequency.  The curve can be fitted using a critical behavior function
$\omega \propto (T_m -T)^\nu $ with $T_m = 0.51$ and $ \nu = 2.2 $.
The characterization of the order of the phase transition is a
difficult issue~\cite{Munoz10,Joyeux_JPCM}. To our knowledge dynamic
exponents for this family of models are not known.

\subsection{Bubble formation}

One of the more interesting aspects of DNA molecule amenable to be
studied in the framework of the PBD mode, is the formation and
stability of the so called DNA bubbles (short open segments of the DNA
chain). The onset of such states have an important role in the
understanding of the operation of DNA molecules.

We observe that the inclusion of a solvation barrier in the model has
a dramatic effect in the dynamics of the system increasing the
lifetime of bubbles. Figure~\ref{fig:bubbles} compares molecular
dynamics numerical simulations for a model with barrier to another
without barrier and at $\rho = 3$. As expected the presence of
barriers modifies the kinetics of base opening and further
closing. The individual trajectories of the different bases clearly
show this behavior. Although the profile of the bubbles is similar in
the two cases, the non-barrier dynamics is completely different from
that with barrier. Without a barrier the opening of the base pair
corresponds to a large amplitude oscillation along the Morse
potential, and an easy closing is favoured. With barrier, the kinetics
is controlled by the presence of two equilibrium states separated by
the solvation barrier. Closing events are more difficult and bubbles
live longer.  Although in this way, we approach to experimental values
of lifetime bubbles (of the order of ns), this fact makes the
simulations more difficult since it is required very long runs to have
good statistics.
 
We see that parameter $\rho$ affects the cooperativity between open
bases and then the bubbles lifetime and width. A large $\rho$ produces
long live bubbles but extremely narrow ones (one or two bases), as
those shown in reference~\cite{SCL_nonl} with $\rho = 25$. With this
value of $\rho$, the transition is extremely narrow and, after a long
transient, bubbles nucleate to drive the unzipping of the whole
chain. A moderate value, $\rho=3$, is good enough to get both longer
and wider bubbles (four to ten bases), as shown in figures.

\begin{figure}[!]
\centering{\includegraphics[width=0.5\textwidth]{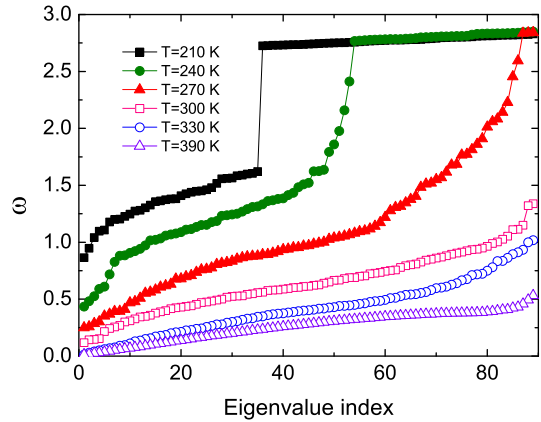}}
\caption{(Color online) PC frequency spectrum at different temperatures for the P5
  promoter sequence. Frequency units are $(D/m)^{1/2}\alpha = 5.15\times10^{12} \rm sec^{-1}$.}
\label{fig:P5} 
\end{figure}

\begin{figure}[!]
\centering{\includegraphics[width=0.5\textwidth]{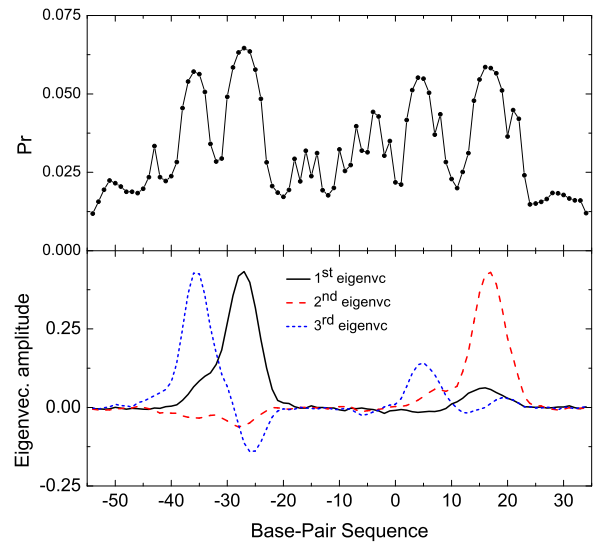}}
\caption{(Color online) Top figure: Probability (not normalised) of opening for
  sequence of P5 promoter. Bottom figure: The three first PC
  eigenvectors at $T= 290 K$.}
\label{fig:P5_comp} 
\end{figure}

\section{The P5 promoter sequence}

\begin{figure*}[]
\centering{\includegraphics[width=0.7\textwidth]{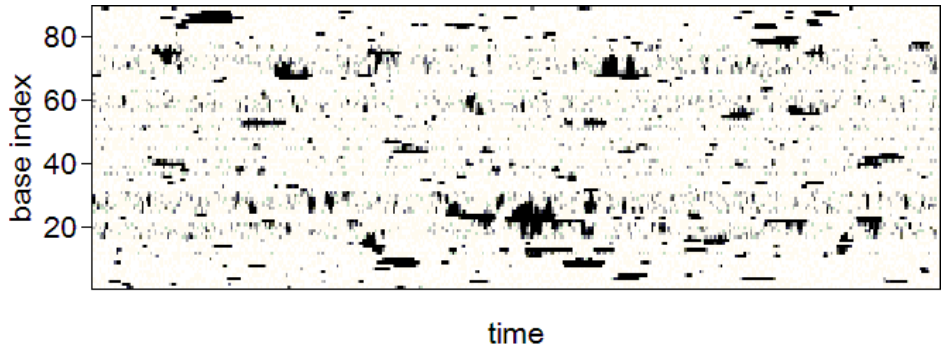}}
\caption{Trajectory of the P5 sequence promoter at $T= 290 K$. The
  larger bubbles appear mainly in the region of the localised PC
  eigenvectors showed in Fig.~\ref{fig:P5_comp}. The time of the
  showed trajectory is 400 ns. }
\label{fig:bubblesP5} 
\end{figure*}

A promoter is a region of the DNA sequence where the transcription to
mRNA is initiated and controlled. In prokaryotic cells, regions of -10
and -35 bp (upstream from the transcription start site) are the most
important sites for transcription regulation.  In this section we
analyse the P5 core promoter which has been widely studied in the
literature~\cite{Kalosakas04, Alexandrov_PLoSCB}.  We show that the
PCA of Langevin trajectories of the molecule clearly identifies the
biologically relevant sites.

The sequence of the P5 promoter is given by the 69 pb:
5'-GTGCCCATTTAGGGTATATATGGCCGAGTGAGCGAGC
AGGATCTCCATTTTGACCGCAAATTTGAACG-3'. We will perform numerical
simulations of this sequence (parameter values for C-G and A-T base
pairs have been given previously). In order to avoid unphysical
denaturation events at low temperature due to finite size
effects~\cite{Peyrard_JPCM}, we apply the following boundary
conditions: First we add 10 C-G bp sequences to the ends of the
promoter to create {\em hard} boundaries. Second, we set the extremes
to 0 (closed) to avoid the complete opening of the chain.

Figure~\ref{fig:P5} shows the evolution of the PC frequency spectrum
with temperature. Several features distinguish these curves from the
homogeneous case. For low temperatures, two bands corresponding to A-T
(lower band) and C-G (upper band) links are clearly identified. The
eigenvectors are localised around rich regions in both kinds of
complementary pairs. At intermediate temperatures the gap between both
bands disappears and the eigenvectors broaden. C-G pairs surrounded by
A-T ones are more likely to open and frequencies diminish.

Close to transition ($ T \approx 345 K $), several modes detach to low
frequencies. The first three modes are strongly localised and show
peaks in four regions (see Fig.~\ref{fig:P5_comp}). These modes
represent the "softest" regions of the sequence, are related with high
probability of opening and drive the unzipping. 

This picture is validated by the study of the three first eigenvectors
(the large eigenvalue or small frequency
ones). Figure~\ref{fig:P5_comp} shows the measured probability of
opening in the sequence, defined by Eq.~\eqref{eq:prob}, and the
computed eigenvectors. An excellent correlation between both figures
is observed. Region +1 corresponds to the transition starting site
(TSS) whereas regions -30 and -40 correspond to the binding sites of
transcription factors i.e. a region rich in A/T pairs like the
TATA-box.  Localised eigenvectors span over regions of ten base pairs
which fairly correspond with the width of the
bubbles.

Figure~\ref{fig:bubblesP5} plots a typical trajectory, showing a few
bubbles mainly at the regions pointed by the PC eigenvectors.  Note
that at this temperature very few opening events occur, so a precise
statistics on bubble formation becomes difficult. However, PC analysis
gives a good account of these sites even with the presence of a few
bubbles. Interestingly, the third eigenvalue shows a clear correlation
between region close to TATA box and the TSS. This represents motion
fluctuations that involve both regions. Thus, modification of one of
them could interfere in the fluctuations in the other opening a
channel for the regulation of transcription.

\section{CONCLUDING REMARKS}

We have studied the dynamics of a DNA molecule using the PBD model
with solvation interaction. Solvation interaction is modeled by
including a barrier in the usual on-site Morse potential. First we
have analyzed the melting transition in an A-T homogeneous chain. We
conclude that the inclusion of the barrier not only modifies the phase
transition but also has a great influence in the dynamics of localised
excitations (bubbles). In combination with the non linear parameter
$\rho$, the width and lifetime of bubbles can be tuned.  We find a set
of parameters which will be suitable for future studies with this
model.

We have applied the model to an inhomogeneous sequence with biological
meaning, a virus promoter.  The study of trajectories using Principal
Component Analysis allows us to detect the biological relevant sites
without the need of long molecular dynamics runs. We have found that
the softest modes of the PC spectrum are highly localised in those
sites. Even more, these sites have been also detected as the most
likely to be opened in bubbles~\cite{Alexandrov_JPCM}. In this work we
cannot elucidate the controversy on whether the DNA can direct its own
transcription as was suggested in~\cite{Kalosakas04} and then make use
of theoretical methods to obtain the functional regions.  However, we
can obtain from the mesoscopic PBD model some regions which are more
sensible to the formation of bubbles (as extracted from the
eigenvectors of the PCA). Of course, the simplicity of the model can
not take into account other effects, like the flexibility of DNA
molecule, but it is able to tackle those related with the sequence.

Finally, we stress the use of a tool, the PCA, to the analysis of
statistical mechanics simulations of simple models to obtain useful
information on linear and nonlinear excitations and the relation with
its phase transition.

\begin{acknowledgments}
We thank L.~M. Flor\'{\i}a and D. Prada-Gracia for helpful comments and discussion. 
Work is supported by the Spanish MICINN project FIS2008-01240, co-financed
by FEDER funds.

\end{acknowledgments}


\begin{thebibliography}{99}
\bibitem{comm1} The addition of the barrier to the standard Morse
  potential that we use, barely modifies the original position and
  curvature of the potential minimun (Eq.~\eqref{eq:Vmod}).
\end{thebibliography}
\end{document}